\documentclass[epj]{svjour}

\usepackage{graphicx,epstopdf}
\usepackage{float}
\usepackage{amsmath}
\usepackage{amssymb}
\usepackage{array}
\usepackage{color}
\usepackage{dcolumn}

\def\ca40{$^{40}\mathrm{Ca}^+$}
\def\T2{$\mathrm{T_2}$}
\def\Pr3{$\mathrm{Pr^{3+}}$}
\def\Eu3{$\mathrm{Eu^{3+}}$}

\newcommand{\PrYSO}[0]{Pr$^{3+}$:Y$_2$SiO$_5\,$}

\def\ket#1{$\left|#1\right>$}
\def\mket#1{\left|#1\right>}
\def\mbra#1{\left<#1\right|}
\def\mrm#1{$\mathrm{#1}$}

\def\eq#1#2{\begin{equation}\label{Eq:#1}#2\end{equation}}
\def\eref#1{(\ref{Eq:#1})}
\def\fref#1{\ref{Fig:#1}}
\def\micro{$\mathrm{\mu}$}

\begin{document}

\title{Fast all-optical nuclear spin echo technique based on EIT}


\author{A.~Walther\inst{1} \and A.~N.~Nilsson\inst{1} \and Q.~Li\inst{1} \and L.~Rippe\inst{1} \and S.~Kr\"oll\inst{1}}
\institute{Department of Physics, Lund University, 221 00 Lund, Sweden, \email{andreas.walther@fysik.lth.se}}

\date{\today}

\abstract{
We demonstrate an all-optical Raman spin echo technique, using Electromagnetically Induced Transparency (EIT) to create the different pulses of the spin echo sequence: initialization, pi-rotation, and readout. The first pulse of the sequence induces coherence directly from a mixed state, and the technique is used to measure the nuclear spin coherence of an inhomogeneously broadened ensemble of rare-earth ions (\Pr3). In contrast to previous experiments it does not require any preparatory hole burning pulse sequences, which greatly shortens the total duration of the sequence. The effect of the different pulses is characterized by quantum state tomography and is compared with simulations. We demonstrate two applications of the technique by using the spin echo sequence to accurately compensate a magnetic field across our sample, and to measure the coherence time at high temperatures up to 11 K, where standard preparation techniques are difficult to implement. We explore the potential of the technique and possible applications.
} 


\maketitle

\section{Introduction}
Quantum technology holds great promise for the future, where several specific applications offer paradigmatic improvements far beyond what is possible using only classical rules. This includes, communication~\cite{Kimble2008}, encryption~\cite{Lo2014}, simulations of other quantum system such as molecules for medicine development~\cite{Georgescu2014} and full quantum computation~\cite{Ladd2010}. To enable quantum aspects of technology it is necessary to control the coherence properties of matter, but today such control are in most cases limited to very restricted environments protecting the sensitive quantum states using cryogenic cooling or ultra high vacuum systems. For practical technological scalability however, it is clearly desirable to search for systems where coherence properties survive in ambient conditions. Very few such systems have been found, mostly related to defects in e.g. diamond~\cite{Balasubramanian2009} or silicon~\cite{Koehl2011}. It is therefore important to search for other systems that also take advantage of special benefits like the ultra-long coherence times of rare-earth ions~\cite{Zhong2015}. It has been known since the 40's that coherence properties of certain nuclear spins survive remarkably well in ambient conditions~\cite{Bloembergen1948}, but finding a good interface with conventional optical technology such as fiber-communications remains an open problem.

Here, we demonstrate an all-optical technique based on Electromagnetically Induced Transparency (EIT), which measures the coherence properties of a long-lived nuclear spin system. Even above liquid Helium temperatures where the optical coherence is very rapidly destroyed, the nuclear spin is shown to remain coherent. The system is initialized from a mixed state by transient EIT pulses with a duration of approximately \mrm{T_{1,opt}/100}. This is much faster than previous similar techniques~\cite{Yale2013} where Coherent Population Trapping (CPT) is used, since CPT require several optical lifetimes for decay. It is shown that the EIT pulses create superposition states directly from mixed states and this is characterized by Quantum State Tomography (QST). For the QST we use an ensemble qubit of rare-earth ions, and it is interesting to note that pulse parameters can be chosen such that the entire inhomogeneous width of the ensemble (170 kHz) is included in the EIT state. A full spin echo sequence of EIT pulses is implemented and demonstrated in two explicit examples: i) cancellation of a dephasing mechanism by compensation of the earth's magnetic field and ii) measurement of the nuclear spin \T2 of \PrYSO for temperatures up to 11 K. The main limitation of the technique is the optical \T2 in relation to the EIT pulse duration. Our particular setup is limited by optical laser power, but we discuss extensions of our scheme to other systems where either higher optical power via pulsed lasers or other atomic species with larger optical oscillator strength can enable all-optical nuclear spin measurements over a broad range of optical \T2's.


\begin{figure}[ht]
	\includegraphics[width=8.5cm,clip=true]{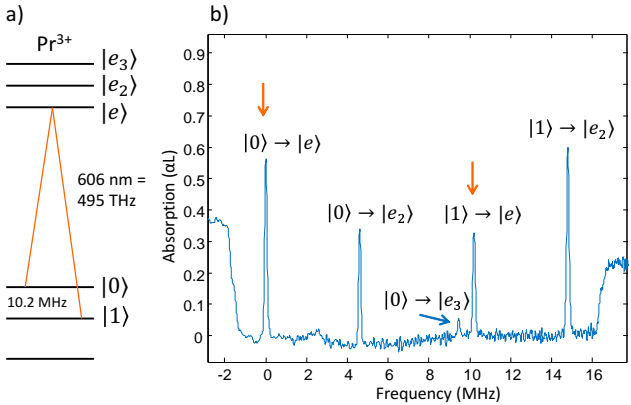}
	\caption{(color online) a) Level structure of the \Pr3 ion. The levels \ket{0}, \ket{1} and \ket{e} form the chosen $\Lambda$-system on which the EIT pulses operate. All transitions used in the population estimation during the experiments are visible in b) and the EIT transitions have been marked with an orange arrow.}
	\label{Fig:levels_abs}
\end{figure}

\section{Theory}
\label{theory}
Consider a \mrm{\Lambda}-type system such as the one depicted in Fig.~\fref{levels_abs} a), where several levels exist but three are chosen for the EIT scheme. The existence of a bright and a dark state will now be briefly derived, but see e.g.~\cite{Fleischauer2005} for a more detailed description. In \Pr3 the two involved transitions, $\mket{0} \rightarrow \mket{e}$ and $\mket{1} \rightarrow \mket{e}$ have different dipole moments, $\mu_{0/1}$, but we can assume that the corresponding laser electric fields, $\mathcal{E}_{0/1}$, can be chosen such that the Rabi frequencies are the same for both transitions, $\Omega = \mu_{0/1} \cdot \mathcal{E}_{0/1} / \hbar$. The interaction part of the Hamiltonian can then be written
\eq{H_int}{
	H_{int} = \hbar \Omega \left( \mket{e}\mbra{0} + \mket{e}\mbra{1} \right) + c.c.
}
We now define an orthogonal basis set for the ground state levels, expressed in the computational one, given by:
\eq{bark_base}{
\left\{ \begin{array}{c}
	\mket{B} = \frac{1}{\sqrt{2}}\left(\mket{0} + \mket{1}\right) \\
	\mket{D} = \frac{1}{\sqrt{2}}\left(\mket{0} - \mket{1}\right)
	\end{array} \right.
}
where the density matrix of each is given by $\rho = \mket{\psi}\mbra{\psi}$. The coupling between each of those ground superposition states and the excited state via the interaction Hamiltonian are now: $\mbra{e}H_{int}\mket{B} = \sqrt{2}\,\hbar \Omega$ and $\mbra{e}H_{int}\mket{D} = 0$. We see that the coupling is increased for the \ket{B} state which is called \emph{bright}, and the coupling is zero for the \ket{D} state which is thus called \emph{dark}. This special type of interaction can be used to create a superposition state between the ground state levels even when they start in a mixed state. The density matrix of a fully mixed state can be written in the bright/dark basis as half of a contribution from each 
\eq{rho_mixed}{
	\rho_{mixed} = 1/2\rho_B + 1/2\rho_D.
}
If a bi-chromatic $\pi$-pulse resonant with the $\Lambda$-levels is now applied to the mixed state, the effect is that the bright component is excited while the dark component is left untouched:
\eq{B_ex}{
	\rho_{total} = 1/2\rho_e + 1/2\rho_D \rightarrow \rho_{ground} = 1/2\rho_D
}
where the last equality is obtained by looking at the ground state levels e.g. with respect to another excited state such as \ket{e_2} (although it does not have trace equal to one and is thus technically not a density matrix, it is illustrative to use it as such in this case). In the experiments presented here this state is sufficient, since the induced coherence is enough for a strong signal from a spin echo sequence. This is a very efficient spin polarization operation as the duration of the full process is only that of the excited state $\pi$-pulse, which can be very short if the Rabi frequency is high enough. For other types of experiments however, one can go one step further, since the excited state population can in principle be detected, e.g. by observing the emission from this level. This can then be used to herald a projection onto either the excited state or $\rho_D$, and in the latter case a pure superposition state have been achieved.

Even though the above descriptions use optical $\pi$-pulses, because the ground states are left in a superposition state, the operation is effectively a $\pi/2$-pulse on the nuclear spin, i.e. it implements the first pulse in the full spin echo sequence (see Fig.~\fref{exp_setup} b). To verify the operation of the sequence, optimize the pulse parameters, and to obtain the scaling for the sequence, a simulation based on the optical Bloch equations was made. A three-level system (see e.g.~\cite{Imamoglu1991}), was simulated numerically, using detunings on both the optical levels and the nuclear spin levels to take the inhomogeneous distributions into account. In Fig.~\fref{QST_tot} a) the path of the two components of a mixed ground state is shown, when an EIT $\pi/2$-pulse and an EIT $\pi$-pulse is applied respectively. A non-traditional EIT $\pi$-pulse is required in the sense that it cannot have the same driving vector as the $\pi/2$-pulse as usual, but needs to rotate the state around a pseudovector which is 90$^{\circ}$ shifted, i.e. it needs to rotate around the state's own center axis. The simulations confirms that this type of $\pi$-rotation works well, even in the case of inhomogeneity on both optical and spin levels.

\section{Experimental setup}
\label{exp_setup}
All experiments have been performed on a \PrYSO crystal with 0.05\% doping concentration, cooled to about 2 K in a liquid bath cryostat. The level-structure of an individual \Pr3 ion is shown in Fig.~\fref{levels_abs} a), where each transition has about 3 kHz homogeneous linewidth, but where the ensemble of \Pr3 ions in the crystal is inhomogeneously broadened to about 5 GHz. In Sec.~\ref{QST}, where the effect of the EIT-pulses is characterized by quantum state tomography, the spectral profile has first been prepared by creating an empty pit containing peaks corresponding to ions in either the \ket{0} or the \ket{1} ground state (the two peaks to the left and the two peaks to the right, respectively, in Fig.~\fref{levels_abs} b)). The width of the peaks is about 170 kHz and such structures can be created using the holeburning techniques described e.g. in~\cite{Rippe2008,Amari2010}. It is ensured that the EIT pulses address the full width of the peaks by selecting a pulse duration of the pulses of 2 \micro s such that the corresponding bandwidth, $\delta f_{bw} \approx 1/\pi t_{dur}$, matches the peak width.

\begin{figure}[ht]
	\includegraphics[width=8.5cm,clip=true]{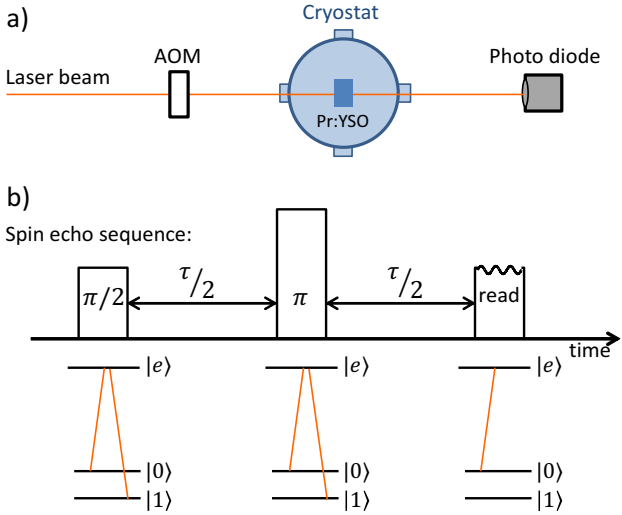}
	\caption{(color online) a) Simplified experimental setup. The bi-chromatic pulses are created in a single pass acousto-optic modulator (AOM). b) The spin echo pulse sequence together with the $\Lambda$-system showing which transitions are active for each pulse. In the readout step, only one of the frequencies is active, and a beating is produced on the difference frequency if the nuclear spin state is still coherent.}
	\label{Fig:exp_setup}
\end{figure}

A simplified experimental setup is shown in Fig.~\fref{exp_setup} a). Here, the laser source at 606 nm (495 THz) is stabilized to a target of 10 Hz and although the coherence created by the EIT pulses does not require a stable laser it is advantageous to ensure that the same ions are always addressed during an experiment. The bi-chromatic pulses contain frequencies separated by the hyperfine splitting in \Pr3 at 10.2 MHz, and are created by a single pass acousto-optic modulator (AOM) with a center frequency of 360 MHz driven by an arbitrary waveform generator with a 1 GS/s sampling rate. The detector is a photo diode connected to a LeCroy HRO 66Zi oscilloscope, with a segmented memory capacity as is utilized in some of the experiments described in the next section. The amount of light deflected by the AOM for a given RF power has been calibrated to the particular laser focus at the crystal sample, allowing each of the bi-chromatic pulse components to be created with an individually specified Rabi frequency. In addition, the optical phase of each component can also be individually set, where the phase difference between the components decide the phase of the created superposition state, which is shown in Fig.~\fref{QST_tot} b). As can be seen in part b) of Fig.~\fref{exp_setup} the $\pi/2$- and $\pi$-pulse in the spin echo sequence are bi-chromatic pulses while the final readout pulse that reveals the spin echo is a single color pulse. If ground state coherence remains in the system at the time, $\tau$, then a readout pulse transfers the component $\mket{0} \rightarrow \mket{e}$, which leads to an excited state coherence at $\mket{1} \rightarrow \mket{e}$. This creates interference with the simultaneous readout pulse and causes a beat note at 10.2 MHz, with an amplitude corresponding to the amount of nuclear spin coherence left after time $\tau$.

\section{Results and discussion}
\label{results}
The results section is divided into four different parts, first the EIT pulses are characterized experimentally, then we present two uses for our technique, and finally we discuss limitations to the technique and scalability to other systems, based on theoretical simulations.

\subsection{Characterization with quantum state tomography}
\label{QST}
In order to show that a single EIT pulse with a duration much shorter than the excited state lifetime can create a pure superposition state from a fully mixed state, we characterize our pulses using quantum state tomography (QST). Even though the final spin echo sequence can be applied without any prior preparation, we use holeburning techniques to prepare isolated peaks inside an otherwise empty spectral pit (as described in ref.~\cite{Rippe2008,Amari2010}), for these characterization experiments. Such peaks are shown in Fig.~\fref{levels_abs} b), where the two large peaks to the left are both transitions from the \ket{0} ground state, and the two rightmost peaks are both transitions from the \ket{1} state. 

The effect of the pulses in the spin echo sequence is simulated, with the results visualized in Fig.~\fref{QST_tot}. If the starting state is fully mixed, then it has equal components from both the two ground states. In the figure, two paths are shown corresponding to the two different ground state components separately, as they are affected by the same pulses. In a) the initialization paths from the spin $\pi/2$-pulse are shown, and the two arrows show the final state after this pulse, which represents the $1/2\rho_D$ state as given by Eq.~\eref{B_ex}. In b) the paths of the components from the initialized state as they are affected by the spin $\pi$-pulse are shown. Keep in mind that the reason why the two different ground states end up in the same final state is because of the involvement of the third level, \ket{e}, through the bi-chromatic pulses, and note that this third level is not shown in the Bloch spheres in Fig.~\fref{QST_tot}.

In order to perform QST we must first be able to estimate the population in each of the available states, visible in Fig.~\fref{levels_abs}. This is done using a least square fitting procedure simultaneous for all states (although, the uppermost excited state, \ket{e_3}, is never involved and is thus assumed to be zero). The full quantum state density matrix can be reconstructed from the population data, corresponding to measurements with three different preceding operations that each reveal the projection onto the three axes of the Bloch sphere:
\eq{rho_exp}{
	\rho = \frac{I + \mathrm{tr}(X\rho)X + \mathrm{tr}(Y\rho)Y + \mathrm{tr}(Z\rho)Z}{2}.
}
The three quantities $\mathrm{tr}(X\rho)$ etc. are obtained experimentally and correspond to the population differences in each of the three measurement cases respectively and $X$, $Y$, $Z$ are the three Pauli matrices (for more details see e.g. Nielsen and Chuang~\cite{Nielsen2000}). Quantum state fidelities can then be obtained by comparing the experimental density matrix to the target ones (e.g. $1/2\rho_D$).

\begin{figure*}[ht]
	\includegraphics[width=18cm,clip=true]{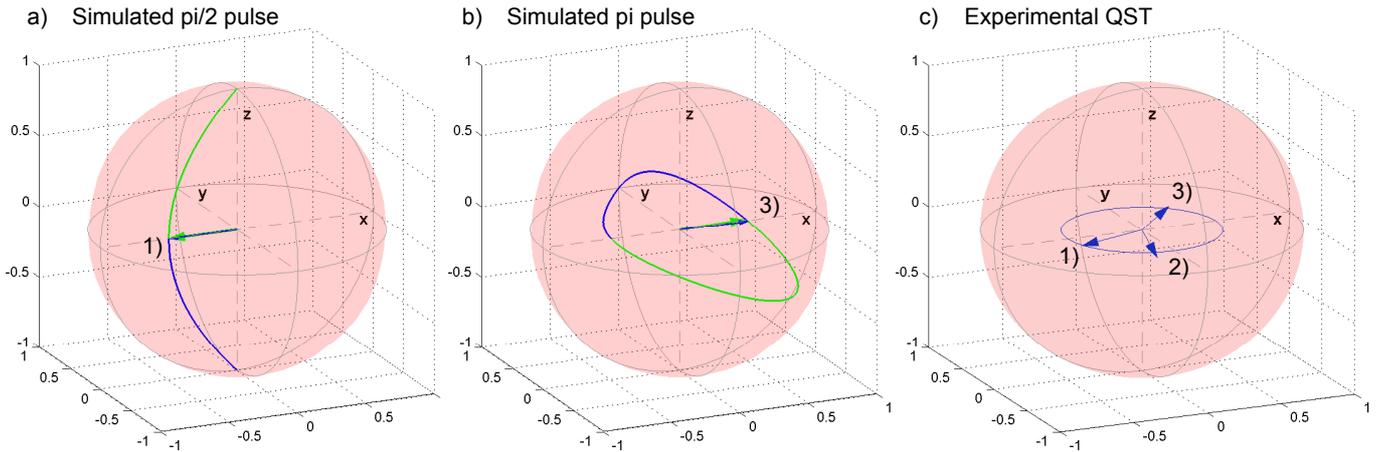}
	\caption{(color online) a) and b) shows optical Bloch simulations of the path for the two principal components (computational basis) of a mixed ground state subject to the bichromatic pulses. In c) the experimental results of the quantum state tomography is displayed, where the blue circle represents the maximum attainable coherence. In both the simulated and the experimental parts the arrows corresponding to case 1) is the state after the $\pi/2$-pulse, case 2) is the state after initialization with a different phase and case 3) shows case 1) after an additional EIT ground state $\pi$-pulse has been applied.}
	\label{Fig:QST_tot}
\end{figure*}

In Fig.~\fref{QST_tot} c) the reconstruction from a QST in three different cases is shown. The three states are 1) EIT initialization to the x-axis, 2) EIT initialization to the y-axis by adding 90$^{\circ}$ to one of the bi-chromatic components, and 3) the initialization as in 1) followed by a rephasing EIT $\pi$-pulse. In the initialization, half of the population is lost to the excited state and if we do not perform any projective detection of this population then the maximum fidelity we can achieve is 75\%. This maximum amount of coherence is represented in the figure as a blue circle with radius 1/2. The qubit fidelities compared to the target states for the three cases are 70\%, 68\% and 67\% respectively, which is remarkable when considering the fact that the qubit is an inhomogeneously broadened ensemble of ions. Each ion has a frequency width of 3 kHz, whereas the whole ensemble is 170 kHz wide. In contrast to many previous EIT experiments where the feature is narrow, we have here chosen spectrally broad EIT pulses, and the high fidelities compared to the target indicate that the EIT effect is indeed applied to most of the population distribution. One can also note that the main loss of fidelity is not due to loss of coherence (arrow length in figure), but rather due to being off the target phase angle, especially for state 3) after the $\pi$-pulse. The reason for this is not certain, but is most likely due to remaining mismatches between the Rabi frequencies of the two components of the EIT pulses. The signal from the spin echo sequence is indifferent to the phase angle of the state however, which means that the amount of coherence created and thus the potential signal strength is therefore even greater than the fidelities imply.

\subsection{Magnetic field compensation}
\label{mag_field}
It has been investigated recently that the presence of small magnetic fields such as the earth's magnetic field can greatly impact the coherence properties of the nuclear spin states~\cite{Heinze2011}. The reason is that even for a small splitting of the hyperfine levels, $\delta f$, distributing the quantum state over different levels will cause them to beat with each other at a timescale given by $1/\delta f$. The EIT spin echo sequence can be used to compensate this splitting in real time, making it a practical lab technique. For \PrYSO, the strongest magnetic g-factor has been measured to be 12 kHz/100\micro T~\cite{Longdell2002}, and although there may exist stray magnetic fields in our cryostat, we believe that in present experiments the earth's magnetic field gives the largest contribution to the splitting. In Lund, where the experiments were carried out, the vertical component of the earth's magnetic field is dominant at about $50$ \micro T, which then gives a splitting of about 6 kHz. The expected nuclear spin \T2 is about 500 \micro s~\cite{Ham1997}, but such a splitting would give a beat minimum after 85 \micro s, drastically limiting the coherence properties.

In order to compensate for magnetic fields in any spatial direction, three pairs of Helmholtz coils with 10 turns each and a radius of 10 cm, was added to the setup. The fast update rate of the experiment depend on two things, the fast EIT sequence, and the ability to use a segmented memory on an oscilloscope. A chain of 30 spin echo sequences with increasing wait time, $\tau$, was pre-programmed into the memory of an arbitrary waveform generator. An oscilloscope with a segmented memory was then used to store only the readout pulses, carrying the resulting beating, in each spin echo sequence. A master trigger was used at the start of the whole chain, followed by secondary triggers at each readout pulse. This enables high temporal resolution capable of resolving the beat note, while at the same time allowing the results from the full chain to be stored simultaneously in the memory. The data was analyzed in real time, extracting the 10.2 MHz beat amplitude using a Fourier algorithm, allowing a full 30 point \T2 decay curve to be displayed at a 1 Hz update rate. This enabled real time tuning of the coil currents in order to compensate for the background magnetic field components in all three directions. The update rate was in our case mostly limited by the oscilloscope master trigger rate, since the spin echo sequence itself plus scramble pulses at the end to reset holeburning effects, requires only 150 ms to run the 30 points in a set.

\begin{figure}[ht]
	\includegraphics[width=8.5cm,clip=true]{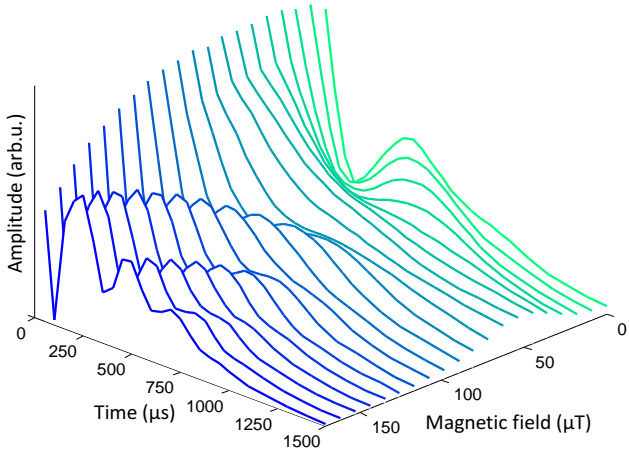}
	\caption{(color online) Coherence decay curves for the nuclear spin vs. the magnetic field in the vertical direction, which was the most sensitive one in our setup. The two other directions were kept at compensated values. The longest (and smoothest) decay occurs at a compensation field of about 50 \micro T, which corresponds to the local vertical component of the earth's magnetic field.}
	\label{Fig:mag_field}
\end{figure}

In Fig.~\fref{mag_field} the result of coherence decay curves at 2 K, for 20 different vertical magnetic field levels is shown, taken while the magnetic field in the two other directions were compensated. The values of the magnetic field axis in the figure is obtained by estimating the field produced by an applied current to our Helmholtz configuration. This estimation gives a compensation in the vertical direction of about 50 \micro T, which matches the earth's magnetic field component in this direction. Off the compensation point, the beatings due to the quantum interference between the split hyperfine states are clearly visible. The oscillation frequency matches the one expected from the sizes of the splittings well, and the fully compensated curve also matches the expected \T2 value well. However, some effects of excitation induced dephasing is observed, and this is discussed in more detail in the next section. 

\subsection{High temperature nuclear spin T$_2$}
\label{high_T}
Above temperatures of approximately 5 K the coherence time, \T2, of the optical transitions decreases very rapidly as a function of temperature, $T_2 \propto 1/T^7$, due to non-resonant two-phonon interactions~\cite{McCumber1963}. It is known that the nuclear spin coherence of \Pr3 does not follow this rapid decay, although the difficulty of preparing a spin polarization when the optical line is broad has limited previous measurements up to only 6 K~\cite{Ham1997}. To extend this we apply a spin echo sequence with our EIT pulses. The combination of EIT interactions being naturally less affected by optical fluctuations and a sensitive Raman beating measurement enables us to obtain signals up to 11 K, corresponding to optical linewidths $(11/6)^7 = $ 70 times broader than previous measurements.

\begin{figure}[ht]
	\includegraphics[width=8.5cm,clip=true]{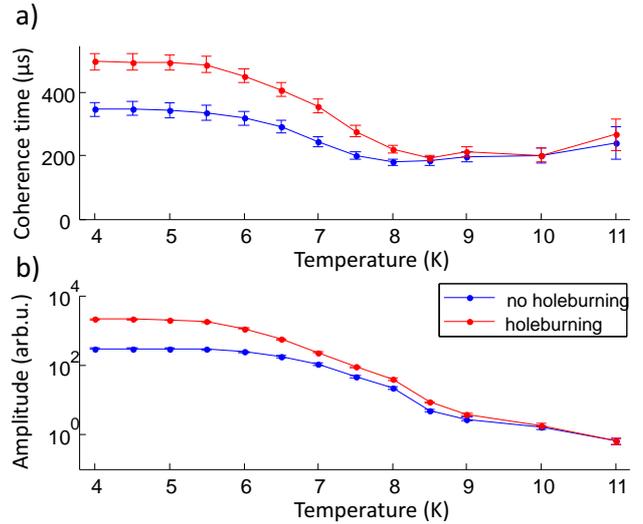}
	\caption{(color online) Coherence time of the nuclear spin as a function of temperature in two cases in a), with and without holeburning before the EIT sequence. The difference between the two curves is discussed in the text. In b) the signal level of the Raman beating is shown as a function of temperature, predictably decreasing as the optical linewidth decreases. For both figures, the error bars correspond to a 95\% confidence interval.}
	\label{Fig:T2_vs_temp}
\end{figure}

Fig.~\fref{T2_vs_temp} shows the coherence time as a function of temperature. Each point in the curves is the \T2 and amplitude respectively, evaluated from an average of 25 decay curves similar to the one at 50 \micro T displayed in Fig.~\fref{mag_field}. From such curves, each with 30 different delay times, $\tau$, exponential decay functions with three different parameters were fitted: amplitude, \T2 and a general offset. The error bars represents the uncertainty of those parameters fitting estimation, and correspond to a 95\% confidence interval.

The lowest point in the figure at 4 K is taken with the sample submerged in a liquid Helium bath, but all other points are taken with cooling through a Helium gas flow otherwise in vacuum. The temperature on the x-axis is the one reported by the cryostat, which has previously been verified by a probe mounted directly at the sample. As can be seen from part b) in the figure, the signal level of the Raman beating decreases strongly with temperature after about 6 K as expected, due to the optical broadening. After about 8 K it is no longer possible to do spectral holeburning and the signal has decreased about two orders of magnitude. Related to this, we see that the \T2 curve in a) displays some curious features, which are now discussed.

The blue curve represents measurements by the EIT spin echo sequence alone on an unprepared sample, while the red curve represents the case when the spin echo sequence was preceded by a preparatory holeburning pulse emptying a 2 MHz absorption region at a frequency corresponding to the $\mket{0} \rightarrow \mket{e}$ transition of the $\Lambda$ system. While this does not affect the action of the EIT sequence itself, it reduces the total amount of ions involved in the experiment, which can remove dephasing caused by other excited ions. Two processes could explain this behavior, either instantaneous spectral diffusion (ISD) from optically excited ions affecting the nuclear spin states, or non-equilibrium resonant phonons. Such phonons are created when the optically excited state decay down to a different, and more probable, Stark level, which then decays down to the ground state by releasing phonons that can be resonantly absorbed by the nuclear spin states destroying their coherence~\cite{Bai1992}. Effects from the standard ISD dephasing can normally be mitigated by reducing the excitation power of the laser, however in this case we have used EIT pulses with a Rabi frequency set to match a $\pi$ pulse area on the optical transition. If this would be a concern, one could avoid these mechanisms e.g. by using a sample with lower doping concentration or go further out to the edge of the inhomogeneous profile, where the same excitation power would excite fewer ions. The behavior after 8 K where the two curves overlap, is due to the fact that when holeburning is no longer possible the same results should be obtained for both sequences. This also indicates that the apparent decrease of the \T2 for higher temperatures is not due to actual decrease of the spin \T2 but rather increased dephasing from larger optical excitation. Therefore, we can conclude that as far up in temperature as we can measure, the nuclear spin is not destroyed by thermal effects but remains coherent.

\subsection{Scaling to other systems}
Even though we have only demonstrated this technique in the particular case of \PrYSO we argue that it can be applied generally over a broad range of systems. In order to determine the limiting factors, optical Bloch simulations for different system parameters were performed. It was found that the signal strength of the Raman beating remains constant until the optical coherence time becomes shorter than the duration of the excitation EIT pulse. In \Eu3 systems e.g., where the starting linewidth is more narrow this occurs at much higher temperatures. From the relation between \T2 and temperature given in ref.~\cite{Koenz2003} it is estimated that it is feasible to reach above 20 K. This would enable the very recent measurements of ref.~\cite{Arcangeli2015} to be implemented all-optically by the method used in the present work.

Outside of rare-earth systems, atomic vapors have also been considered for quantum memories, and these systems often involve optical excitations that are dipole allowed, giving much shorter excited state lifetimes in the order of tens of ns, but at the same time having much higher oscillator strengths, allowing $\pi$-pulses with a much shorter duration. This enables our method to work for systems with a broad range of \T2 times, as we will now analyze in more detail.

Quite generally, the Rabi frequency is directly proportional to the transition dipole moment, $\mu$, which in turn causes the duration of a $\pi$-pulse to be inversely proportional to it: $T_{\pi} \propto 1/\mu$. At the same time, \T2 of a transition is upper limited by 2\mrm{T_1}, which is inversely proportional to the square of the transition dipole moment, i.e.: $T_2 \propto 1/\mu^2$ (see e.g. \cite{Hilborn1982}). Combining these two relations gives
\eq{T_scaling}{
	T_{\pi} \propto \sqrt{T_2}
}
If we take as a simple assumption that the achievable laser intensity is roughly constant for different time scales, then the signal level of the EIT sequence can be simulated for different time scales given the constraints of Eq.~\eref{T_scaling}. In Fig.~\fref{scaling}, we have plotted the achievable nuclear spin fidelity when limited by a laser intensity similar to the one in our experiments, which is approximately 100 mW focused to an area of about 70 \micro m. The simulations show that the cutoff in this figure for lower times occur when the optical \T2 becomes shorter than the duration of the EIT pulses, where the duration is set by the desire to reach a $\pi$ pulse area using the available intensity.

\begin{figure}[ht]
	\includegraphics[width=8.5cm,clip=true]{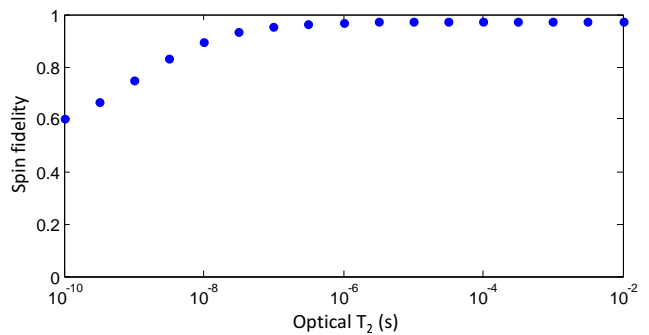}
	\caption{Fidelity of the nuclear spin that is left at the end of the EIT spin echo sequence as a function of the optical \T2 of the system from simulations. Since any amount of coherence is enough to achieve a signal from a spin echo sequence, the result shows viability of the method down to at least 100 ps.}
	\label{Fig:scaling}
\end{figure}

While high fidelity spin polarization is achievable as long as the optical \T2 times is above 10 ns, obtaining a signal from a spin echo sequence can be achieved as long as there is some amount of coherence induced in the system. This allows the the EIT spin echo method to be used down to about 100 ps, which covers most if not all, system that could be interesting for quantum information purposes. 

However, keep in mind that this result contains the simple restriction of constant laser intensity. When targeting system with short lifetimes, pulsed lasers are often available that can deliver much higher peak intensities, which will further extend the scheme. A special case of this may be when short pulsed lasers are used to target $\Lambda$-systems where the laser is not bi-chromatic but two of the levels are covered by the inherently broad (but still phase coherent) spectrum of the pulse. In this case, an EIT superposition can also be initiated~\cite{Berman2005} and used as a coherence probe. Although, the readout mechanism would have to be different than the beating used here, e.g. a rotation onto the population of the states.

\section{Conclusions}
\label{conclusions}
We have described and experimentally demonstrated a fast and efficient all-optical spin echo technique to generate a coherent ground state spin superposition from mixed state. The method is based on EIT and is particularly useful e.g. in higher temperature settings or when short and fast sequences are needed. The created dark superposition states have been characterized by quantum state tomography and then applied to two usefulness of the technique has been demonstrated for two selected cases. Firstly, an accurate and real time compensation of the background magnetic field components in all three directions. Secondly, the nuclear spin of \T2 of \Pr3 was measured by all-optical means up to a temperature of 11 K. Comparisons to simulations show this to be limited by the optical power of the laser. Possibilities to further extend this by means of e.g. pulsed lasers were discussed. The method can be used in a wide variety of systems, since even though the optical \T2 must be longer than the pulse duration, the useful regime was shown to be achievable in most cases.

\begin{acknowledgement}
This work was supported by the Swedish Research Council, the Knut \& Alice Wallenberg Foundation. The research leading to these results also received funding from the People Programme (Marie Curie Actions) of the European Union's Seventh Framework Programme FP7 (2007-2013) under REA grant agreement no. 287252 (CIPRIS) and Lund Laser Center (LLC).
\end{acknowledgement}

\bibliographystyle{unsrt}
\bibliography{EIT_init}

\end{document}